# Electron scattering and capture rates in quantum wells by emission of hybrid optical phonons

V. N. Stavrou, C. R. Bennett*, O. M. M. Al-Dossary† and M. Babiker*.

Institute of Technical Physics, DLR, Pfaffenwaldring 38-40, Stuttgart, Germany.

*Department of Physics, University of York, Heslington, York YO10 5DD, UK.

†Department of Physics, King Saud University, Riyadh, Saudi Arabia.

**Abstract**

Intra and intersubband scattering rates and electron capture rates are considered when mediated by hybrid optical phonons in an AlAs/GaAs/AlAs double heterostructure confined between two outer metallic barriers. In evaluating scattering rates we concentrate first on an infinite quantum well for the electrons and show that the presence of the outer metal barriers results in reductions of the intra and intersubband scattering rates due to the suppression of the interface-like modes. For a quantum well (QW) with a finite depth we find that the outer barriers are responsible for the existence of a discrete energy spectrum above the well. The electron capture process under these circumstances is defined as the electron transition from the first electron subband above the well to all possible subbands inside the well by the emission of hybrid phonons. Explicit calculations reveal that the capture rates are characterized by sharp peaks, referred to as electron resonances which arise when a new electron state is generated on increasing the quantum well width. Other sharp peaks are identified as phonon resonances and arise when the energy of the initial state differs by a phonon energy from an electron state at the bottom of a quantum well subband.

## 1. Introduction

The interactions of electrons with the optical phonons in quantum wells (QWs) and superlattices are of crucial importance in the physics of semiconductor heterostructures. In particular, the intrasubband and intersubband scattering rates for electrons and holes by emission of optical phonons have special significance for studying the optical and electrical properties in such structures. It has thus been long realised that the accurate modelling of devices, such as hot-electron transistors, quantum well laser structures and infrared detectors requires these scattering rates to be well-described [1].

There are a number of models used for the description of optical phonons in a layered system [1,2]. We could assume that the bulk phonons are still applicable, however, it is known that optical phonons become confined in such systems. Two models that can be used to describe this confinement are the dielectric continuum (DC) model and the hybrid model. Interestingly the total scattering rates are normally consistent whichever model is used in a standard QW [1] but it is of interest to investigate if this is the case in more complicated structures and whether the simpler DC model can still be assumed. However, it should be remembered that the hybrid model does describe the phonons in a more complete manner as it includes dispersion and gives the correct symmetry for the modes [1,2]; for effects due to individual phonon states, for example Raman scattering, the DC model does not produce the correct results.

The first calculation of intra and intersubband scattering rates by emission of hybridons was produced by Constantinou and Ridley [1], assuming that the barrier region of the heterostructure played no role in determining the scattering rates by hybrid modes and so taking into account electron scattering only with modes of frequencies in the *reststahl* band of the QW material. Subsequently, Ridley used phonon tunnelling [2] to evaluate the scattering rates including effects due to the barrier region with the scattering process involving the emission of triple hybridons. One of the aims of such calculations was to examine the suggestion that scattering rates could be reduced by a suitable choice of the heterostructure materials and dimensions. There are two different methods mentioned in the literature identified as possible means for a desirable reduction of the scattering rates. First, as in the work of Pozela et al [3,4], the dielectric continuum (DC) model was used to examine this possibility when a foreign monolayer was inserted in the middle of the quantum well. Also, the work by Ridley [5] uses the hybrid model (again with a foreign monolayer inserted). These models yield different results. However, even though the monolayer changes the optical phonon modes, it has been shown [6] that an infinitesimally thin monolayer should not modify the scattering rate.

The second method identified for a possible reduction of the scattering rates is by confining the double heterostructure between two metallic barriers [7]. In this paper, we concentrate on this second method by considering a quantum well structure where the interfaces between the well and barrier material are at $z = \pm L$ and there are outer metal barriers for $|z| > D$. We evaluate the intrasubband and intersubband scattering rates by the emission of optical phonons as described by the hybrid model [8]. Earlier work on a similar situation was done by Constantinou [7] who considered the emission of DC optical phonons.

Finally we evaluate the electron capture rates by the emission of hybrid phonons and compare the results with those predicted by the emission of DC optical phonons [9]. The capture rate is defined as the electron emission rate from the bottom of the subband with an energy just above the barrier energy into all possible subbands within the quantum well mediated by either DC or hybrid optical phonons. In order to calculate these rates we assume that the structure is confined between infinite barriers. This ensures that the energy spectrum above the well is not a continuum. It is only in this way that it is possible to identify a discrete initial electron state and evaluate the change of the energy of this state with quantum well width. We also show that the non-parabolicity of the electron subbands has little effect on the capture rates [10].

**2. Double hybrid phonons.**

A hybrid phonon consists of three kinds of mode, TO/LO/IP (transverse optical, longitudinal optical and interface polariton modes) [8], and is called a triple hybrid. The TO part of the hybrid mode does not have any electric field and so it does not contribute to the Fröhlich coupling mechanism [1] and, hence, can be discarded for the purpose of electron-phonon interactions. This approximation of the triple hybrid model is called the double hybrid model.

The boundary conditions for the hybrid model are electromagnetic and mechanical [8]. Long-wavelength lattice vibrations both acoustic and optical, can be adequately described by the continuum approximation arising from the macroscopic theory of lattice vibrations. The mechanical conditions demand that the normal components of the stress tensor be continuous. This can be shown to amount to the continuity of the components of the ionic displacement field $\mathbf{u}$ which are tangential and perpendicular to any interface [11].

Consider the hybrid modes appropriate for the layered structure described earlier. The precise details of the ionic displacement fields and the corresponding electric potential field $\Phi$ depend on the mechanical boundary conditions we choose [1,2,8]. We discuss below two possibilities, case (a) and case (b), approximating the mechanical boundary conditions for the AlAs/GaAs heterostructure, namely:

(a) $\mathbf{u} \neq \mathbf{0}$ (well region) $\mathbf{u} \approx \mathbf{0}$ (barrier region).

(b) $\mathbf{u} \approx \mathbf{0}$ (well region) $\mathbf{u} \neq \mathbf{0}$ (barrier region).

Even when $\mathbf{u} \approx \mathbf{0}$ there is still an electric field associated with the IP type modes through the structure.

It is important to note the influence of bulk dispersion, which is a primary ingredient of the hybrid model when applied to double heterostructures. In heterostructures made with AlN/GaN, the reststrahl bands overlap in the first Brillouin zone and so the modes that propagate in the quantum well may also propagate in the barrier region [11]. Therefore, there is need to take into account the mode propagation in both regions and, as a result, the approximate mechanical boundary conditions in (a) and (b) above are not applicable for the nitrides. On the other hand, in heterostructures made with AlAs/GaAs the reststrahl bands do not overlap [1,2] and, as a result, we may ignore the modes in one region while the modes in the other region are propagating. Therefore, the conditions in (a) and (b) are more suitable when considering an AlAs/GaAs heterostructure.

### 2.1. Hybridon mode functions in the well

Considering case (a) first, namely that in the barrier region $|z| > L$ the ionic displacement vanishes $\mathbf{u}_{QW} = 0$, the ionic displacement field in the quantum well region $|z| < L$ for the symmetric (S) modes can be written in terms of LO part (subscript L) and interface part (subscript I),

$$\mathbf{u}_{QW}^{S} = \left[ A_L^S \cos(k_{L1} z) + A_I^S \cosh(q_{||} z), \quad 0, \quad A_L^S \frac{ik_{L1}}{q_{||}} \sin(k_{L1} z) - A_I^S i \sinh(q_{||} z) \right], \qquad (1)$$

where we have omitted the common factor $e^{i(q_{||} x - \omega t)}$, $\mathbf{q}_{||}$ is the in-plane wavevector and the confinement wave vector $k_{L1}$ for the LO part is given by

$$\omega^2 = \omega_{L1}^2 - v_{L1}^2\left(k_{L1}^2 + q_{||}^2\right), \tag{2}$$

where $v_{L1}$ is the dispersive velocity of the LO mode in material 1. For the antisymmetric (A) modes the ionic displacement field in the quantum well region is similarly found to be

$$\mathbf{u}_{QW}^A = \left[ A_L^A \sin(k_{L1}z) + A_I^A \sinh(q_{||}z), \quad 0, \quad -A_L^A \frac{ik_{L1}}{q_{||}}\cos(k_{L1}z) - A_I^A i\cosh(q_{||}z) \right]. \tag{3}$$

where $A_L^{S,A}$ and $A_I^{S,A}$ are coefficients to be determined.

The corresponding electric field in the QW region (material 1) is given by $\mathbf{E_L} = -\tilde{e}_1\mathbf{u_L}$ for the LO part and $\mathbf{E_{IP}} = -\tilde{e}_1 s_1(\omega)\mathbf{u_{IP}}$ for the interface part where $\tilde{e}_1 = \sqrt{\rho(\omega_{L1}^2 - \omega_{T1}^2)/\varepsilon_0\varepsilon_\infty}$ , $s_1(\omega) = \left(\omega^2 - \omega_{T1}^2\right)/\left(\omega_{L1}^2 - \omega_{T1}^2\right)$ and $\rho$ the reduced mass density for the ions in material 1 [12]. The boundary conditions at the outer metallic barriers are such that the tangential part of the electric field is zero. In the barrier region of the double heterostructure there exists an electric field associated with the IP modes even if it is assumed that the ionic displacement vanishes in that region. Therefore, in the barrier region $L < |z| < D$, the electric field corresponding to the symmetric ionic displacement is given by

$$\mathrm{E}_{QW}^S = \tilde{e}_1 F^S \left[\sinh\left(q_{||}\left(D - |z|\right)\right), \quad 0, \quad -i\,\mathrm{sgn}(z)\cosh\left(q_{||}\left(D - |z|\right)\right)\right], \tag{4}$$

and that associated with the antisymmetric ionic displacement is given by

$$\mathrm{E}_{QW}^A = \tilde{e}_1 F^A \left[\mathrm{sgn}(z)\sinh\left(q_{||}\left(D - |z|\right)\right), \quad 0, \quad -i\cosh\left(q_{||}\left(D - |z|\right)\right)\right], \tag{5}$$

where $F^{S,A}$ are coefficients to be determined by applying the normalisation condition and the boundary conditions, namely $u_z = 0$ and $E_x$ and $D_z$ continuous at $z = \pm L$. The electric field already satisfies the condition that its tangential component vanishes at the outer interfaces with the metal which is already manifest in Eqs. (4) and (5).

In the non-retarded limit, the scalar potentials for the symmetric and antisymmetric modes can be written as

$$\phi_{QW}^{S} = i\frac{\tilde{e}_1}{q_{||}}\begin{cases}\left(A_L^S \cos(k_{L1}z) + s_1 A_I^S \cosh(q_{||}z)\right), & |z| < L, \\ -F^S \sinh\left(q_{||}\left(D-|z|\right)\right), & L < |z| < D,\end{cases} \tag{6}$$

$$\phi_{QW}^{A} = i\frac{\tilde{e}_1}{q_{||}}\begin{cases}\left(A_L^A \sin(k_{L1}z) + s_1 A_I^A \sinh(q_{||}z)\right), & |z| < L, \\ -F^A \operatorname{sgn}(z)\sinh\left(q_{||}\left(D-|z|\right)\right), & L < |z| < D,\end{cases} \tag{7}$$

Once one applies the boundary conditions, the dispersion relation for the symmetric modes emerges in the form

$$\tanh(q_{||}L) + s_1\frac{k_{L1}}{q_{||}}\tan(k_{L1}L)\left[1+\frac{\varepsilon_1}{\varepsilon_2}\tanh(q_{||}b)\tanh(q_{||}L)\right] = 0\,, \tag{8}$$

where $b = D - L$. Similarly, for the antisymmetric modes we obtain the dispersion relation

$$\coth(q_{||}L) - s_1\frac{k_{L1}}{q_{||}}\cot((k_{L1}L)\left[1+\frac{\varepsilon_1}{\varepsilon_2}\tanh(q_{||}b)\coth(q_{||}L)\right] = 0\,. \tag{9}$$

Eqs. (8) and (9), are applicable in the case where the ionic displacement field is localised in the quantum well region corresponding to case (a) mentioned above. They are very similar to those found in [1] and the expression inside the square brackets when equal to zero gives the DC interface mode dispersion relation [13]. The dispersion curves corresponding to Eqs. (8) and (9) are shown in the Figure 1 for large and small barrier widths. For a large barrier width, branches of different symmetry cross while branches of the same symmetry anti-cross, as is clear in Figure 1(a). A similar quantative feature of hybrid modes has been pointed out in earlier work by Constantinou and Ridley [1] who dealt with the different problem of a quantum well surrounded by infinite half spaces of the barrier material.

The hybridons have the behaviour of a mixture of confined and interface polariton modes. The interface polariton modes are responsible for the crossing of the hybridon branches of different symmetry. For small in-plane wavevector the IP dispersion curves start at large values of frequency for the antisymmetric mode (or small values for the symmetric mode) in the GaAs reststrahl band and for large in-plane wavevectors they become flat. The confined phonons have a fixed transverse wavevector and, hence, follow the parabolic dispersion of the LO mode (which appears flat for the scale shown on the in Figure 1). The hybrid mode branches change as they take on the character of confined and IP modes for different frequencies.

The role of interface modes in the hybridon behaviour is more obvious in the case of small barrier widths. For this case the confined mode dispersion is unaffected. As shown in Figure 1(b), the symmetric and antisymmetric branches do not cross for frequencies in the vicinity of the longitudinal optical frequency of the well material. The antisymmetric IP mode for small barrier widths has almost no dispersion and, as a result, the higher frequency antisymmetric hybridons take on this character. Of crucial importance for the hybridon behavior at small barrier width is the symmetric IP mode. As illustrated in Figure 1(b), the symmetric interface mode in the GaAs reststrahl band changes rapidly for small in-plane wavevectors and becomes flat at large wavevectors. Thus, the symmetric hybrid modes also have a higher dispersion in this region.

The Hamiltonian for a hybrid mode is given by [1]

$$\hat{H} = \frac{M}{2V_c}\left[\int \hat{\dot{u}}^2(\mathbf{r})d^3\mathbf{r} + \omega^2 \int \hat{u}^2(\mathbf{r})d^3\mathbf{r}\right], \qquad (10)$$

where $\hat{\mathbf{u}}(\mathbf{r})$ is the ionic displacement operator for the hybridons and $V_c$ is the crystal volume. Using this Hamiltonian we obtain via the canonical quantization procedure the final coefficient. Since the normalization only depends on the displacement field, it is unaffected by the metallic barriers [1]. We now consider the hybrid modes due to the barrier regions corresponding to condition (b).

### 2.2. Hybrid mode functions in the barrier regions

The ionic displacement field in the barrier region $L < |z| < D$ corresponding to condition (b) can be derived by following a similar procedure. We have

$$\begin{aligned}\mathbf{u}_{BW}^{S} = \Big[ & q_{\|}\left(A_1^S \cos(k_{L2}z) + A_2^S \sin(k_{L2}z)\right) + iq_{\|}\left(A_3^S e^{-q_{\|}z} + A_4^S e^{q_{\|}z}\right), \quad 0, \\ & k_{L2}\left((A_1^S \cos(k_{L2}z) + A_2^S \sin(k_{L2}z)\right) - q_{\|}\left(A_3^S e^{-q_{\|}z} + A_4^S e^{q_{\|}z}\right)\Big].\end{aligned} \tag{11}$$

The antisymmetric mode displacement field, $\mathbf{u}_{BW}^{A}$, is defined in an analogous manner but with a minus sign when $z$ is negative. Once again, the mechanical displacement in the well region, $|z| < L$, is zero but there is, nevertheless, an electric field associated with the IP modes in the quantum well which is given by

$$\mathbf{E}_{BW}^{S} = \tilde{e}_2\left[A_I^S \cosh(q_{\|}z), \quad 0, \quad -iA_I^S \sinh(q_{\|}z)\right], \tag{12}$$

$$\mathbf{E}_{BW}^{A} = \tilde{e}_2\left[A_I^A \sinh(q_{\|}z), \quad 0, \quad -iA_I^A \cosh(q_{\|}z)\right]. \tag{13}$$

The corresponding electric potential turns out to be as follows

$$\phi_{BW}^{S} = \frac{i\tilde{e}_2}{q_{\|}}\begin{cases} -A_I^S \cosh(q_{\|}z), & |z| < L, \\ A_1^S \cos(k_{L2}z) + A_2^S \sin(k_{L2}z) + s_2\left(A_3^S e^{-q_{\|}z} + A_4^S e^{q_{\|}z}\right), & L < |z| < D, \end{cases} \tag{14}$$

for the symmetric barrier modes and

$$\phi_{BW}^{A}=\frac{i\tilde{e}_2}{q_{\|}}\begin{cases}-A_I^A\sinh(q_{\|}z), & |z|<L,\\ \mathrm{sgn}(z)\left[A_1^A\cos(k_{L2}z)+A_2^A\sin(k_{L2}z)+s_2\left(A_3^Ae^{-q_{\|}z}+A_4^Ae^{q_{\|}z}\right)\right], & L<|z|<D,\end{cases}\tag{15}$$

for the antisymmetric barrier modes. By applying the boundary conditions at the interfaces we are able to determine all unknown coefficients except one (which is determined by the normalization condition) and we also deduce the dispersion relations as

$$\begin{aligned}&\left(\frac{s_2k_{L2}}{q_{\|}}\right)^2\sin(k_{L2}b)\left[\frac{\varepsilon_1}{\varepsilon_2}\sinh(q_{\|}b)+\cosh(q_{\|}b)\,\mathrm{ct}(q_{\|}L)\right]-\\ &\frac{s_2k_{L2}}{q_{\|}}\left[2\frac{\varepsilon_1}{\varepsilon_2}-\cos(k_{L2}b)\left(2\frac{\varepsilon_1}{\varepsilon_2}\cosh(q_{\|}b)+\sinh(q_{\|}b)\,\mathrm{ct}(q_{\|}L)\right)\right]-\\ &\frac{\varepsilon_1}{\varepsilon_2}\sinh(q_{\|}b)\sin(k_{L2}b)=0,\end{aligned}\tag{16}$$

where ct($x$) is coth($x$) for symmetric modes and tanh($x$) for antisymmetric modes. For brevity we do not display the corresponding expression for the final coefficient.

The hybridon dispersion curves are shown in Figure 2 for both the symmetric and antisymmetric modes. It can be seen that the number of the hybridon branches in the barrier region is quite large. The important factor determining the number of hybrid branches in the barrier region is clearly the barrier width; as this increases, more branches appear.

As we mentioned earlier, the presence of the symmetric IP contribution, shown in Figure 2(a), is responsible for the curvature of the symmetric hybridon branches before they become flat. For frequencies close to the TO frequency of the AlAs reststrahl band, the IP mode has no effect on the curvature. In the case of the antisymmetric hybridons illustrated in Figure 2(b), the IP antisymmetric branch is flat and, as a result, the frequency does not change rapidly with the in-plane wavevector.

### 3. Intersubband and Intrasubband scattering rates

Let us consider first the simplest case in which the electrons are subject to an infinite quantum well. The electrons are then confined totally in the quantum well $|z| \le L$. Thus, the electron wavefunction vanishes in the barrier region $L \le |z| \le D$ and, as a result, there is no contribution to the scattering rates from the barriers.

Our first task is to evaluate intersubband scattering, where electrons from the first excited state (n=2) make downward transitions into the ground state (n=1), and intrasubband scattering, whereby electrons scatter within the ground state (n=1). The symmetric and antisymmetric solutions of Schrödinger's equation are well-known:

$$\begin{aligned} \Psi_1(\mathbf{r}) &= \frac{1}{2\pi\sqrt{L}} e^{i\mathbf{k}_{\|}\cdot\mathbf{r}_{\|}} \cos\left(\frac{\pi}{2L} z\right), \\ \Psi_2(\mathbf{r}) &= \frac{1}{2\pi\sqrt{L}} e^{i\mathbf{k}_{\|}\cdot\mathbf{r}_{\|}} \sin\left(\frac{\pi}{L} z\right). \end{aligned} \tag{17}$$

The scattering rate by emission of hybrid optical phonons is given by Fermi's golden rule with the interaction Hamiltonian taken as $e\Phi$, where $\Phi$ is the quantised Coulomb field incorporating all possible modes, Eqs (6), (7), (14) and (15). In order to compare our predictions to those arising from the DC model [7], we consider the variation of the intersubband scattering rate with barrier width for scatters from the bottom of the second subband into states in the lowest subband. This is shown in Figure 3 for a fixed well width $L = 20$Å. As is clear from this figure the contribution from the well region is non-zero at $D - L = 0$ and it increases with the barrier width until a point where it then becomes flat. The non-zero value at $D - L = 0$ is due to the LO part of the well modes since in this limit there are no interface-like modes. Note, however, that the symmetry of these modes is opposite to those of the confined modes in the DC model, although the total rate is the same [1]. Increasing the barrier width is accompanied by more branches emerging but their individual contributions to the rates becomes progressively smaller for larger barrier widths. As a result, the scattering rates become flat for large barrier width. For the DC model the increase is due to the stronger GaAs and weaker AlAs interface modes, which become less suppressed as the outer barriers are moved further away from the quantum well. The hybrid phonons also take on more interface-like character as $D$ is increased.

The variations of the intrasubband scattering rateswith barrier width ($D$ - $L$) for a fixed $L$ are presented in Figure 4. It can be seen that the intrasubband scattering rates increase with $D$ - $L$ for small barrier width and become constant at large values. In a similar manner to Figure 3, the LO and interface parts are responsible for this behaviour. However, since it is the stronger AlAs interface mode which interacts with the electrons in this case (due to symmetry) the reduction is greater when ($D$ - $L$) is small for both the DC and hybrid phonon results.

The fact that the results emerging from the two models (DC and hybrid) are the same is attributed to an approximate sum-rule [13,14] involving the use of bulk, DC and hybrid phonons, which should hold irrespective of how the electron system is described. The sum-rule stresses the fact that although the optical modes predicted by different models are different (both in frequency and form), the total scattering rate from each model will have approximately the same value. In other words, no matter what method is used to describe the phonons, completeness ensures practically the same result for the total scattering rate [14]. The small bulk dispersion in GaAs and AlAs, in particular, is responsible for ensuring that the results are very close.

## 4. Electron capture rates

We have seen that the hybridons described in Section 2 for large in-plane wavevectors behaved like DC phonons [8]. One expects then that the capture rates by the emission of hybrid optical phonons to be comparable to those arising by emission of DC phonons. In Section 2 we analysed the dispersion and functional form of the hybrid modes which are characterized as having (a) the ionic displacement non-zero in the well region and vanishing in the barrier regions and (b) the ionic displacement non-zero in the barrier regions and vanishing in the well region. In this section we evaluate the capture rates for these two sets of phonon modes and the total capture rates for the GaAs/AlAs double heterostructure quantum well. The electron capture mechanism is defined as the transition from the bottom of the first discrete electron state [9] above the well to all possible states into the well by the emission of optical phonons as given by Fermi's golden rule. It is convenient to present capture rates in terms of the characteristic rate $\Gamma_0$ for bulk GaAs. We have $\Gamma_0 = e^2 / 4\pi\varepsilon_0\hbar \left(1/\varepsilon_{\infty 1} - 1/\varepsilon_{s1}\right)\sqrt{2m_1^*\omega_{L1}/\hbar}$ which is $8.7\times10^{12}\mathrm{s}^{-1}$ using typical parameter values [1].

Figure 5 shows the variation of the capture rate with *L*, calculated for electrons at an energy at the bottom of the first subband above the barrier energy. In Figure 5(a) we show the contribution to the capture rate from the quantum well region arising from the emission of DC phonons and of hybrid modes. The capture rate by emission of DC phonons presented in Figure 5(a) arises from the contribution of the confined DC modes in the quantum well region and the interface polariton DC modes in the GaAs reststrahl band. This is because these DC phonons have similar frequencies to the hybrid phonons of the well material and are the types of mode which couple in the hybrid theory to generate the hybrid modes due to the well. In Figure 5(b) we show the contribution to capture rates from the barrier regions by emission of DC and hybrid phonons. The capture rate by emission of DC phonons presented in Figure 5(b) is due to the confined modes in the barrier regions and the DC interface polaritons of the AlAs reststrahl band. The total capture rates by the emission of DC and hybrid phonons are shown in Figure 5(c). The rates show regular peaks with an overall approximately linear increase as $L$ increases.

The peaks are associated with two distinct physical characteristics of the system: the onset of electron resonances and threshold emission of phonons. The electron resonances, which occur when a new subband enters the well from the energy region above the well, are distinguished by a sharp drop, while the phonon peaks, which occur when the energy difference between the initial state and a subband in the well equals a phonon energy, have sudden thresholds as $L$ increases. Since the point at which a subband enters the well and the phonon modes also depend on the value of $D$, these resonances will depend on the position of the outer metal barriers. However, this dependence is weak.

The electron and phonon resonances appear for similar well widths and the magnitudes differ only slightly. This is expected since the values of $L$ at which the electron resonances occur are only dependent on the material system, not the phonons, and the frequencies of the DC and hybrid phonons are very close (because bulk dispersion has little effect) so the phonon resonances occur at similar values of $L$. Once more, the fact that the magnitude is the same for the two models is attributed to the approximate sum-rule [13,14] between bulk, DC and hybrid phonons, which should hold in these circumstances.

## 5. Conclusions

In order to calculate the intrasubband and intersubband transition rates by the emission of hybrid modes for an AlAs/GaAs quantum well we have adopted the approximate picture that mechanical oscillations in the barrier and well regions are independent. The modes that are propagating in the well (barrier) region with the appropriate frequency are the only non-vanishing ionic displacements. In the barrier (well) region these modes have vanishing amplitudes ionic displacements. This approximation is suitable for the case of an AlAs/GaAs quantum well because there is no overlap in the reststrahl frequency bands in the first Brillouin zone. In the case of nitride heterostructures this approximation **does not** work since there is an overlap in the reststrahl frequency bands [15].

The transition rates of electrons in a quantum well with outer metal barriers involving the emission of DC phonons are in close agreement with those predicted by the emission of hybrid modes. This confirms that bulk dispersion does not affect the reduction which was obtained due to the suppression of the interface modes by the introduction of the outer metal barriers.

The capture rates were found to be strongly dependent on the well width and at specific well widths they show sharp peaks. A subset of these resonances can be explained as due to a subband energy difference becoming equal to a phonon energy. Hence these peaks are referred to as *phonon resonances*. The second set of peaks arises as new electron subbands enter the well as the width increases. The appearance of these resonances is associated with an increased probability distribution inside the well and an increase of the matrix element. Hence these resonances are called *electron resonances*. The two resonance sets appear in pairs and the general features are predicted by both the hybrid and DC optical phonon models. The hybridons and DC phonons give rise to the same magnitudes for the phonon resonances and this feature is expected because of the weak effect of bulk dispersion and the approximate sum-rule in operation.

## 6. Acknowledgements

The authors would like to thank Prof. Brian Ridley and Dr. Nick Zakhleniuk for useful discussions.

**Figure captions**

**Figure 1.** The dispersion curves of hybridons for a GaAs/AlAs quantum well with $L$ = 14Å and (a) $D$ = 300Å or (b) $D$ = 28Å. The full curves correspond to DC interface polariton modes; the dashed curves to the symmetric modes and the dotted curves to the antisymmetric modes. The indices (S) and (A) denote the symmetry of the IP modes.

**Figure 2.** The dispersion curves of (a) the symmetric and (b) the antisymmetric hybrid modes in the barrier region for a GaAs/AlAs quantum well with $L$ = 14Å and $D$ = 28Å where the full curve corresponds to the DC interface polariton mode of the same symmetry and the dotted curves are the hybrid modes.

**Figure 3.** The intersubband scattering rates for an electron in an initial state at the bottom of the first excited state scattering into the ground state in an AlAs/GaAs/AlAs quantum well system with $L$ = 20Å against the barrier width $D$ - $L$. The total rate due to hybrid phonons is the solid curve while the total rate due to DC phonons is the dotted curved. The dashed and chained curves are the contribution from the hybrid phonons due to the well and barrier materials, respectively.

**Figure 4.** The intrasubband scattering rates for an electron in the ground state with initial energy $E_{||} = 2\hbar\omega_{L1}$ in an AlAs/GaAs/AlAs quantum well at fixed well width $L = 20\overset{0}{A}$ against the barrier width $D$ - $L$.. The total rate due to hybrid phonons is the solid curve while the total rate due to DC phonons is the dotted curved. The dashed and chained curves are the contribution from the hybrid phonons due to the well and barrier materials, respectively.

**Figure 5**. The electron capture rate from the first state above the well into all states in the well against $L$ for a GaAs/AlAs system with $D = 300$Å via emission of (a) well modes, (b) barrier modes and (c) all modes using the hybrid model (solid curve) and DC model (dotted curve) for the optical phonons.

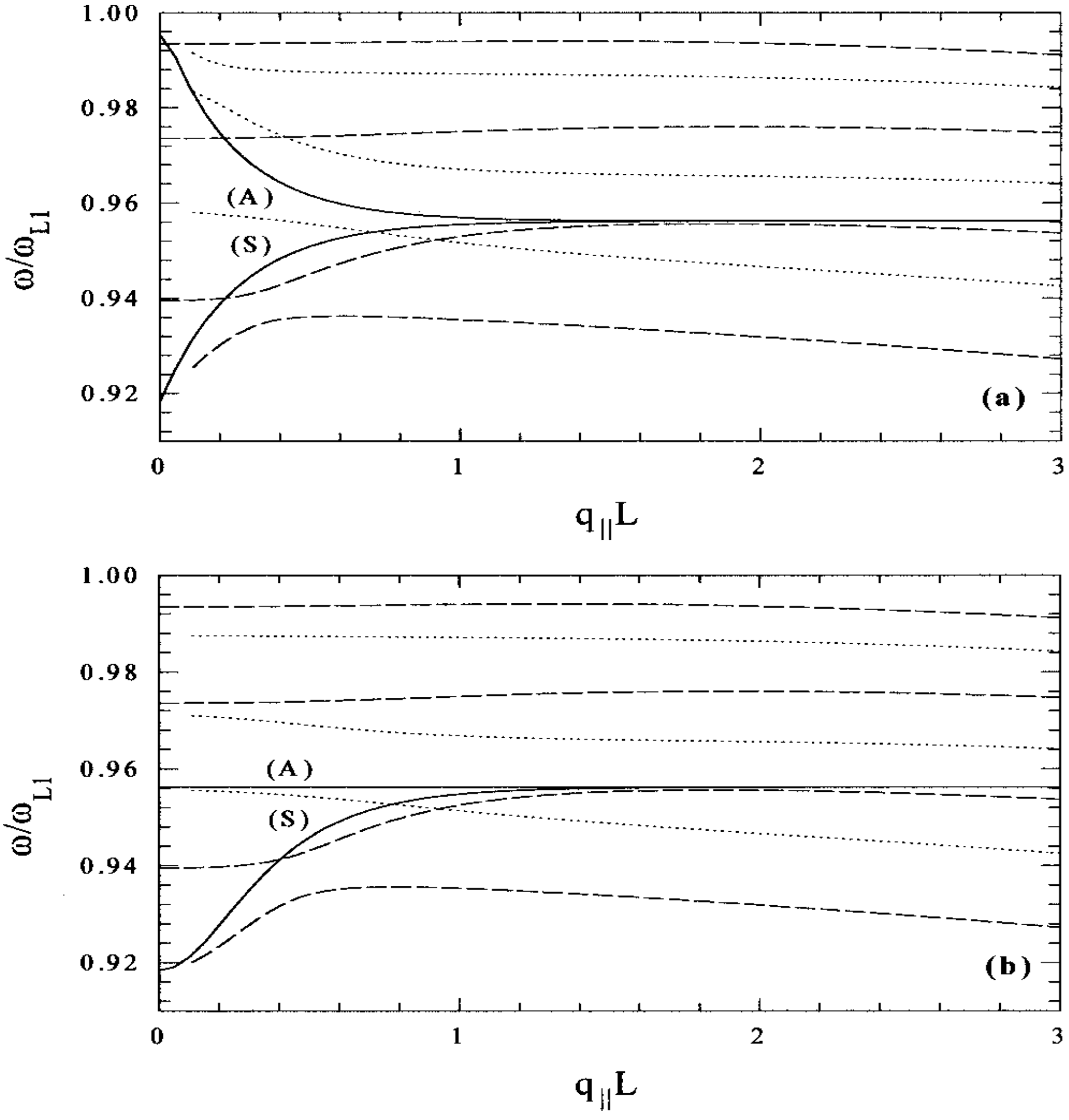

**Figure 1.** V. N. Stavrou

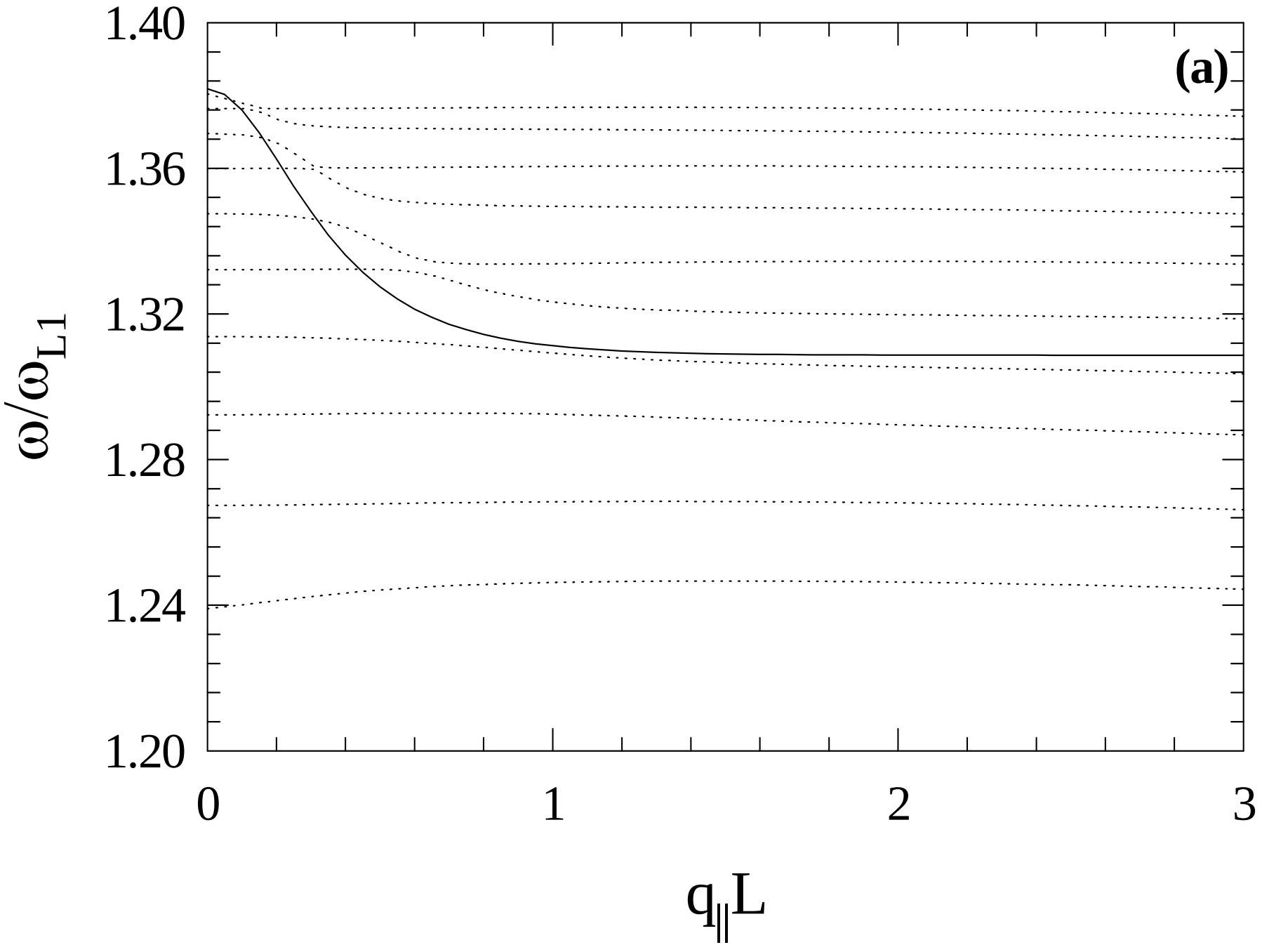

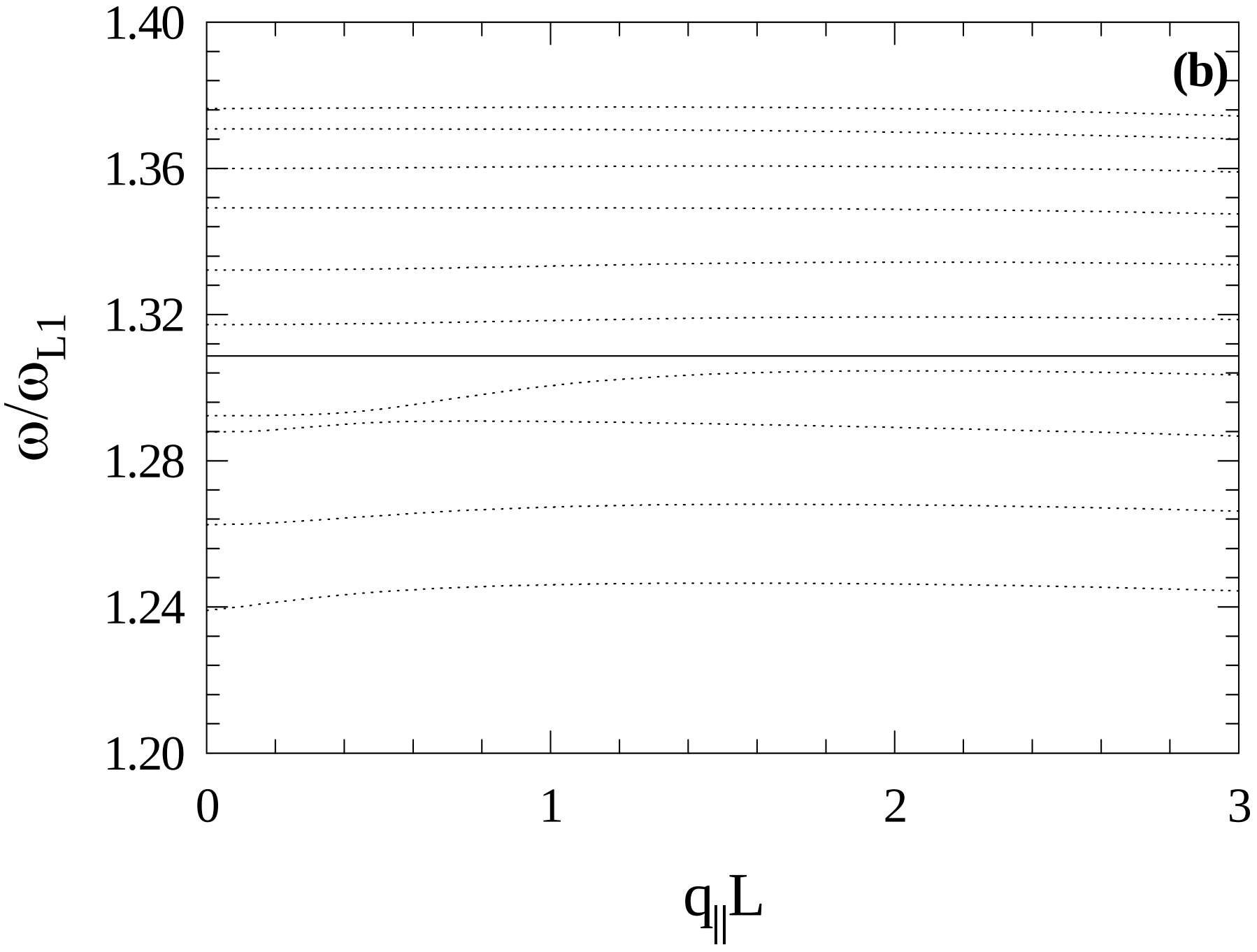

**Figure 2.** V. N. Stavrou

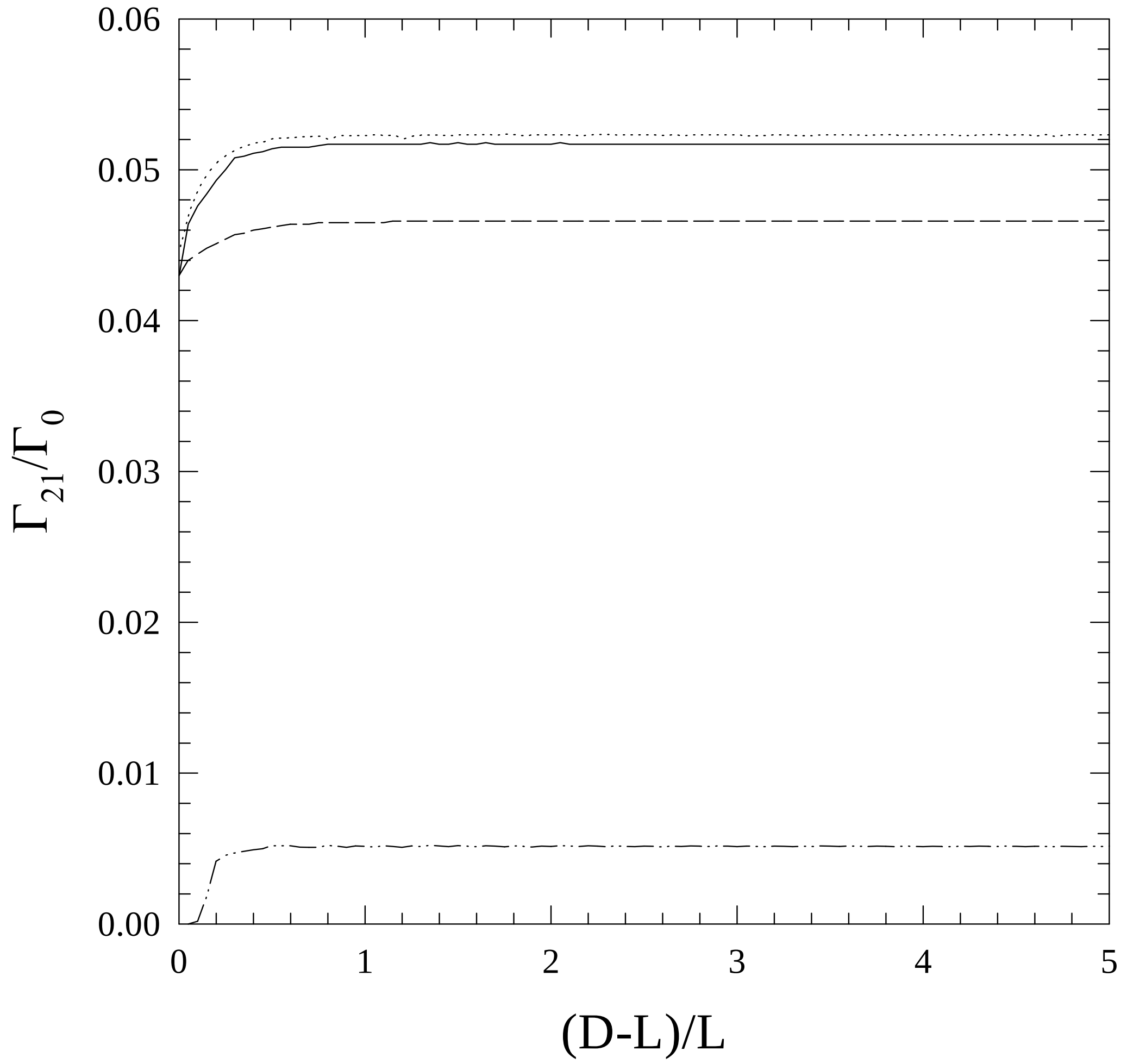

**Figure 3.** V. N. Stavrou

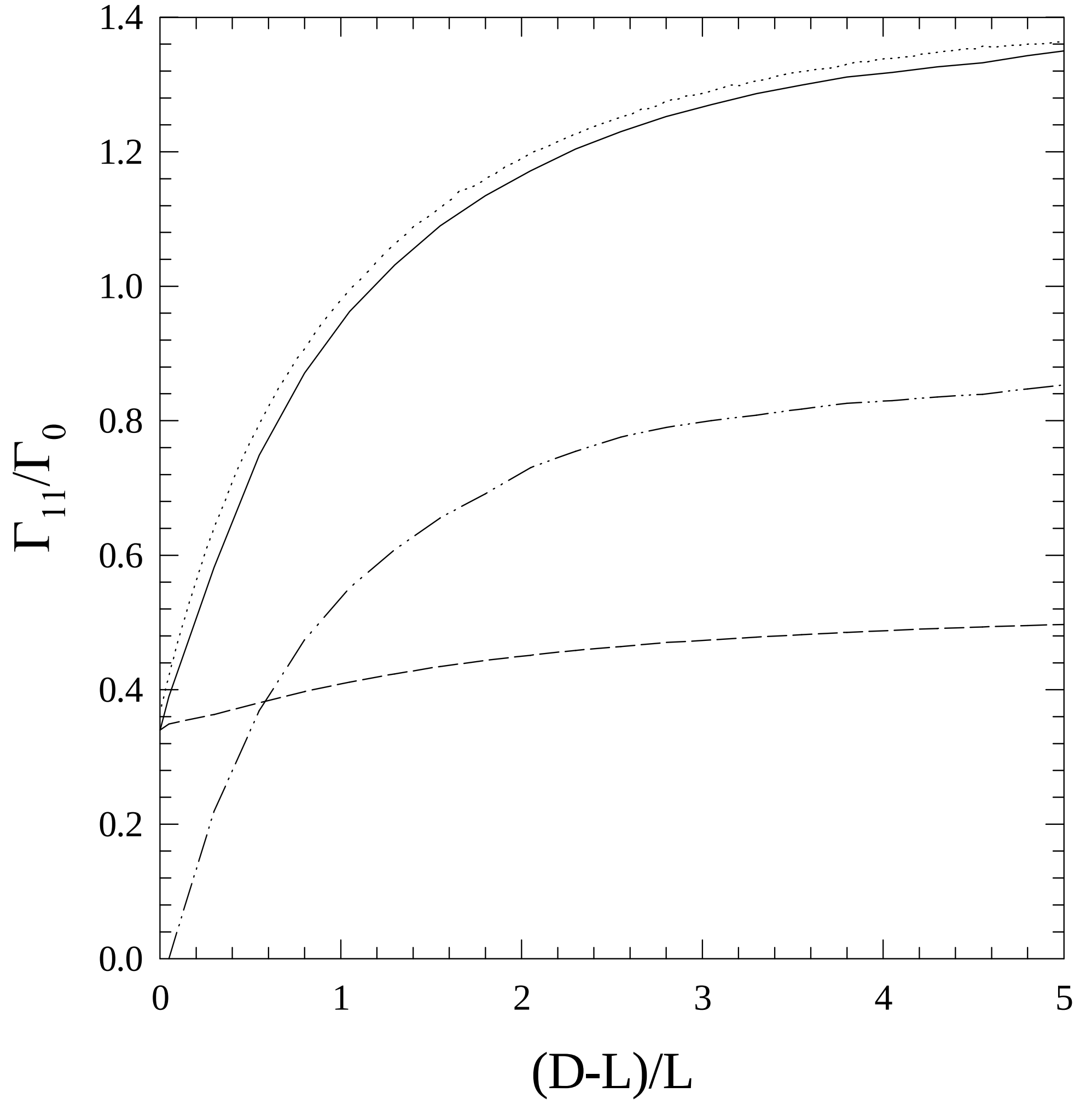

**Figure 4.** V. N. Stavrou

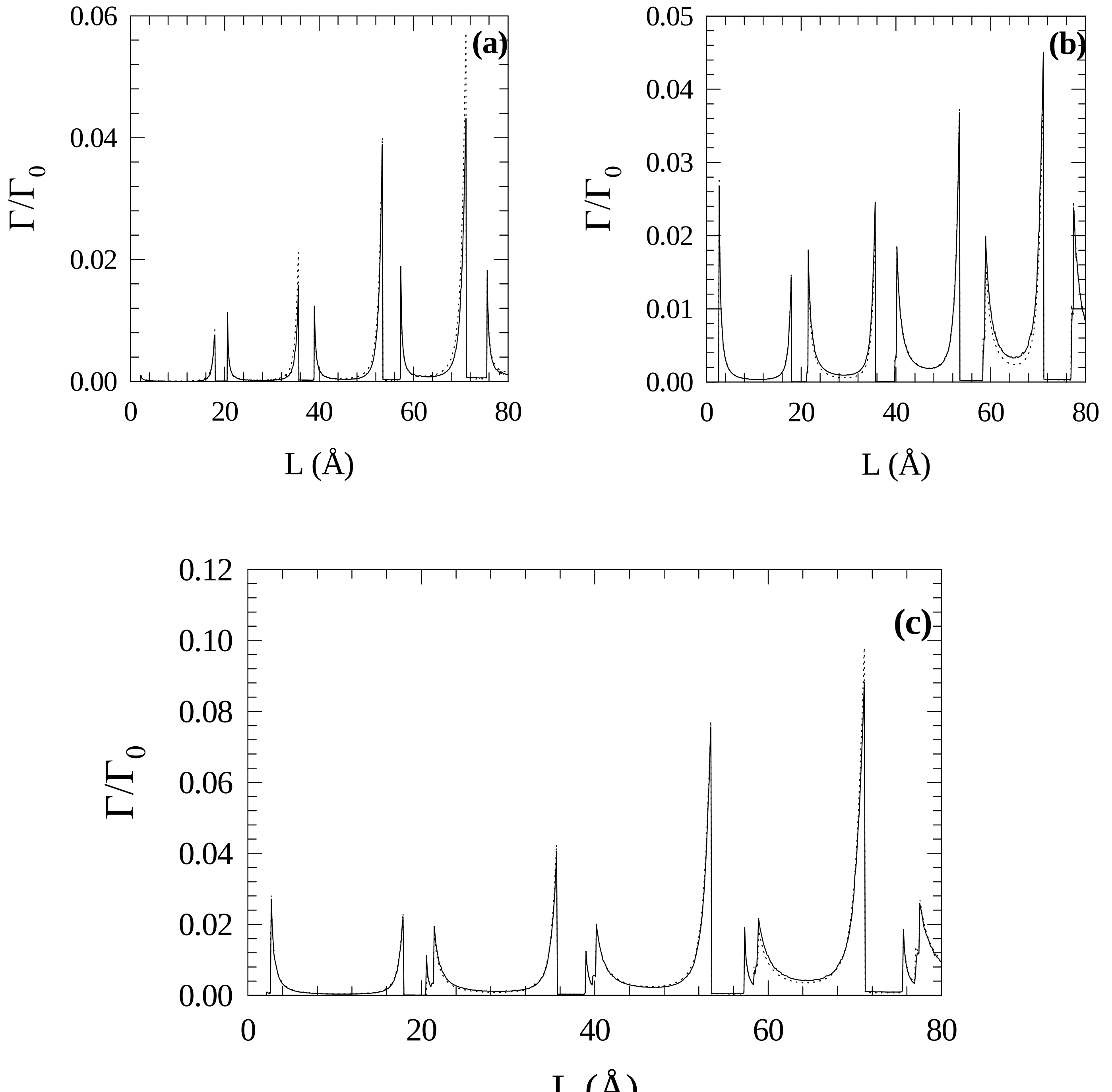

**Figure 5.** V. N. Stavrou